\documentclass{PoS}

\title{Towards the determination of the photon parton distribution
  function constrained by LHC data}

\ShortTitle{Towards the determination of the photon PDF constrained by
  LHC data}

\author{\speaker{Stefano Carrazza}\thanks{On behalf of the NNPDF Collaboration.}\\
  Dipartimento di Fisica, Universit\`a di Milano \& INFN, Sezione di Milano\\
  E-mail: \email{stefano.carrazza@mi.infn.it}}

\abstract{ We provide a discussion of the impact of a subset of
  Drell-Yan data from LHC on the determination of the photon parton
  distribution function (PDF), using the NNPDF methodology. In
  previous work we have shown that the photon PDF determined from
  deep-inelastic scattering (DIS) data only has large uncertainties,
  suggesting the need for more data from other processes such as
  Drell-Yan, which unlike DIS, includes photon-induced contributions
  at leading order in QED. We describe the inclusion of ATLAS
  Drell-Yan $W$, $Z/\gamma^{*}$ data, which is a subset of the LHC
  data used in a final photon PDF determination, by means of a
  reweighting procedure. We show the impact of such data by comparing
  the reweighted photon PDF with the photon PDF from DIS, highlighting
  the reduction of uncertainties at medium/small-$x$. We conclude that
  the Drell-Yan data from LHC allows a reasonably accurate
  determination of the photon PDF.}

\FullConference{XXI International Workshop on Deep-Inelastic Scattering and Related Subject - DIS2013,\\
  22-26 April 2013\\
  Marseille, France}

\begin{document}

\section{Introduction}

In Ref.~\cite{Carrazza:2013wua} we have shown, using the latest NNPDF
technology~\cite{Ball:2012cx}, that the photon parton distribution
function (PDF) obtained from deep inelastic scattering (DIS) data has
large uncertainties. Furthermore, by computing with {\tt HORACE}
~\cite{CarloniCalame:2007cd} the $Z\rightarrow\mu^{+}\mu^{-}$
production in proton-proton collision, we have confirmed that the
photon PDF from DIS is not sufficiently precise for the level of
accuracy required by the LHC physics.

The large uncertainties of the photon PDF from DIS are due to the lack
of photon-induced contributions to DIS processes, when including QED
corrections up to leading order (LO) in $\alpha$ (i.e. $O(\alpha)$) to
PDFs. A possible cure for this issue is the inclusion of Drell-Yan
data where there are photon-induced contributions at $O(\alpha)$: for
example in $Z/\gamma^{*}$ production the $\gamma\gamma \rightarrow
l^{+} l^{-}$ process.

Here we present the determination of the photon PDF, for a PDF set
which is determined at NLO in QCD and LO in QED, by including a subset
of Drell-Yan data from LHC. Currently, fast interface codes for QCD
computations do not include contributions from photon PDFs, so we
decided to include these data through a reweighting procedure. In
short, the reweighting assigns weights to PDF replicas which satisfies
the new data by computing the $\chi^{2}$ between the new experimental
data and theoretical predictions obtained from each replica. Technical
details about the reweighting and unweighting procedure are presented
in Ref.~\cite{Ball:2010gb,Ball:2011gg} and will not be discussed here.

\section{Reweighting with ATLAS data}

We start the reweighting procedure by creating a prior PDF set with
500 replicas, combining the photon PDF from DIS with five copies of
the NNPDF2.3 NLO set~\cite{Ball:2012cx} at the initial scale
$Q^{2}_{0}=2$ GeV$^{2}$ and then evolving with the combined QCD+QED
DGLAP evolution equations. The prior PDF set thus defined preserves
the prediction accuracy of a global set for pure QCD computations:
indeed we have explicitly verified that PDF sets with and without QED
corrections are statistically equivalent which means that the
corrections to the quark and gluon due to their mixing with the photon
are indistinguishable from statistical fluctuations. Also, the
violation of the momentum sum rule due to the inclusion of
the photon PDF is less than 1\%.

We perform reweighting with two LHC data sets: ATLAS $W$,
$Z/\gamma^{*}$ rapidity distributions~\cite{Aad:2011dm} and ATLAS
Drell-Yan high mass differential
cross-section~\cite{Aad:2013iua}. These data sets provide constraints
to the photon PDF at medium and large-$x$ thanks to the presence of
photon-induced contributions in the $Z/\gamma^{*}$ and $W$
production. The QCD computation is performed with {\tt
  APPLGRID}~\cite{Carli:2010rw} for ATLAS $W$, $Z/\gamma^{*}$ rapidity
distributions data and {\tt DYNNLO}~\cite{Catani:2009sm} for ATLAS
Drell-Yan high mass data, while, in both cases we have computed the
Born and NLO photon-induced contributions with {\tt
  HORACE}~\cite{CarloniCalame:2007cd}, knowing that NLO contributions
have a very small effect.

\begin{figure}
  \begin{centering}
    \includegraphics[scale=0.4]{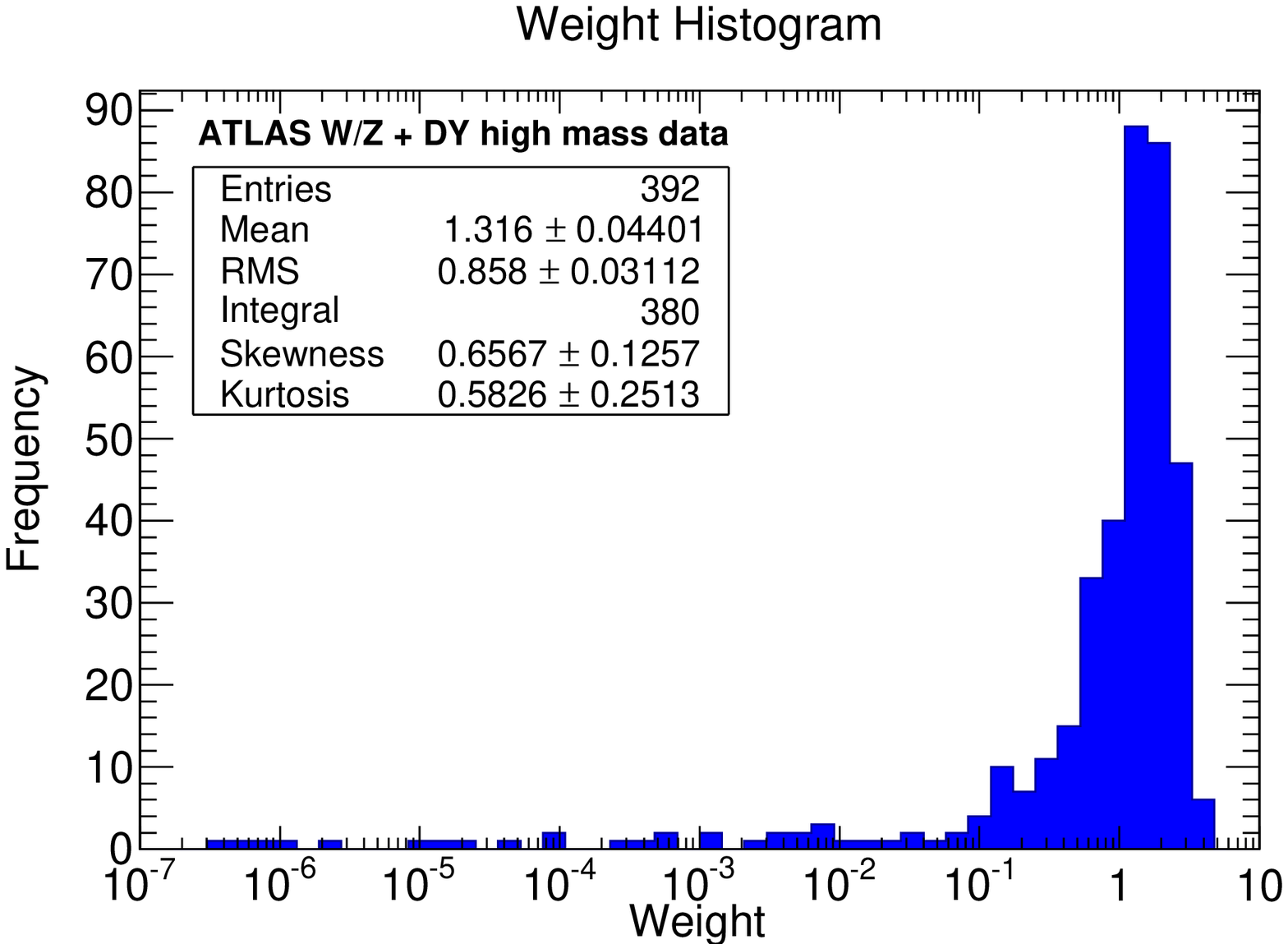}\includegraphics[scale=0.4]{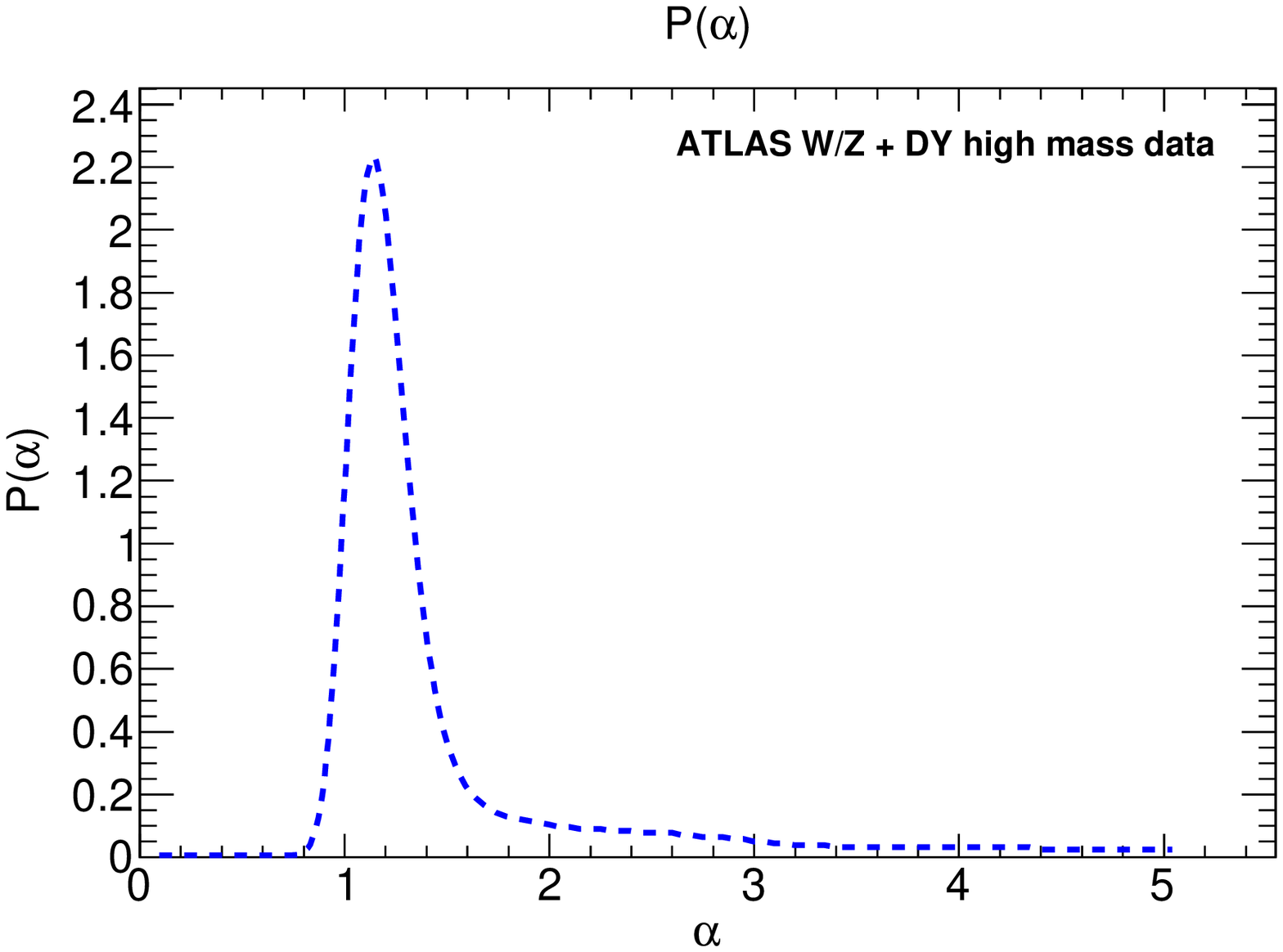}
    \par\end{centering}
  \caption{\label{fig:weights} On the left, the distribution of
    weights after performing the reweighting with ATLAS $W$,
    $Z/\gamma^{*}$ rapidity data and ATLAS Drell-Yan high mass
    data. The prior set of $N_{\rm rep}=500$ replicas after the
    reweighting procedure is reduced to $N_{\rm eff}=266$ effective
    replicas. On the right, the associated $P(\alpha)$ distribution.}
\end{figure}

Figure~\ref{fig:weights} shows the weights and $P(\alpha)$
distributions, defined in~\cite{Ball:2010gb}, for the present
reweighting. Recall that $\alpha$ is the rescaling of the covariance
matrix: $\alpha=1$ means that uncertainties are correctly
estimated. From these distributions we conclude that the procedure is
consistent and that the data is compatible with the reweighted PDF
set. After performing the reweighting we obtain a total number of
effective replicas of $N_{\rm eff}=266$ and the $\chi^{2}$ to the
ATLAS data is reduced from 1.98 to 1.14. In Table~\ref{table:chi2} we
collect the $\chi^{2}$ and $P(\alpha)$ values for the individual
reweighting procedure, looking at the $\chi^{2}$ values before and
after reweighting we confirm that data far from the $W$,
$Z/\gamma^{*}$ peaks, where the photon contribution is substantial,
contributes strongly to constrain the photon PDF.

%%%%%%%%%%%%%%%%%%%%%%%%%%%
\begin{table}
  \centering
  \begin{tabular}{c||c||c|c}
    \hline
    NLO &  ATLAS all  & ATLAS $W$, $Z/\gamma^{*}$  & ATLAS DY high mass  \\ 
    \hline
    \hline
    $\chi^2_{\rm in}$  & 1.98 & 1.20 & 3.78 \\
    \hline
    $\chi^2_{\rm rw}$  &  1.14 &  1.15 & 1.01 \\
    \hline
    $N_{\rm eff}$  & 266 & 364 & 326 \\
    \hline
    $\langle \alpha \rangle$  & 1.48 & 1.24 & 1.53 \\
    \hline
  \end{tabular}
  \caption{\label{table:chi2} Reweighting properties: $\chi^{2}_{\rm in}$ and $\chi^{2}_{\rm rw}$ are before and after reweighting, respectively; $N_{\rm eff}$ is the number of effective replicas; $\langle \alpha \rangle$ is the average rescaling of the covariance matrix.}
\end{table}
%%%%%%%%%%%%%%%%%%%%%%%%%% 

%%%%%%%%%%%%%%%%%%%%%%%%%%%%%%%%%%%%                                   
%%%%%%%%%%%%%%%%%%%%%%%%%%%%%%%%%%%%

\begin{figure}
  \begin{centering}
    \includegraphics[scale=0.39]{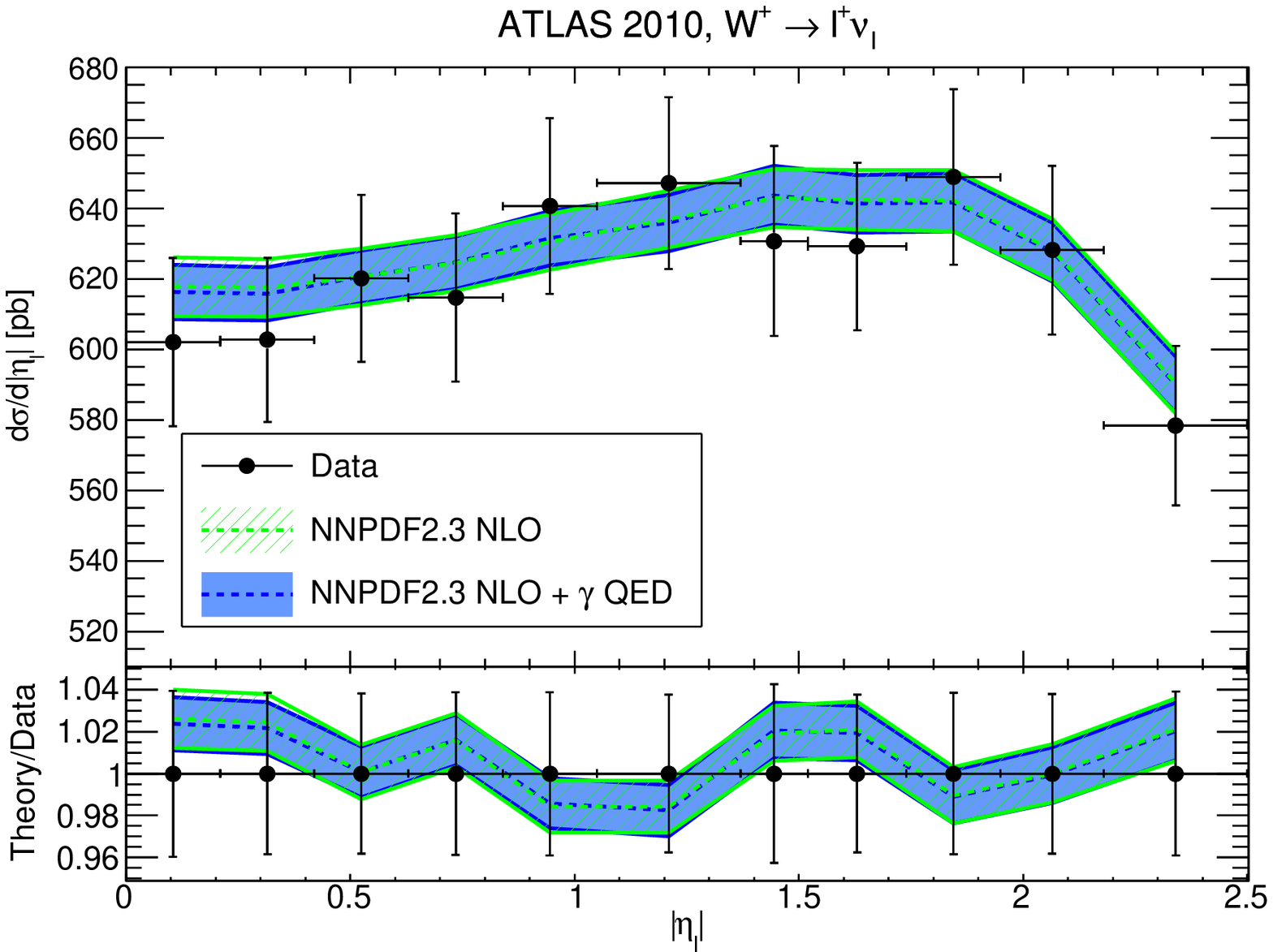}\includegraphics[scale=0.39]{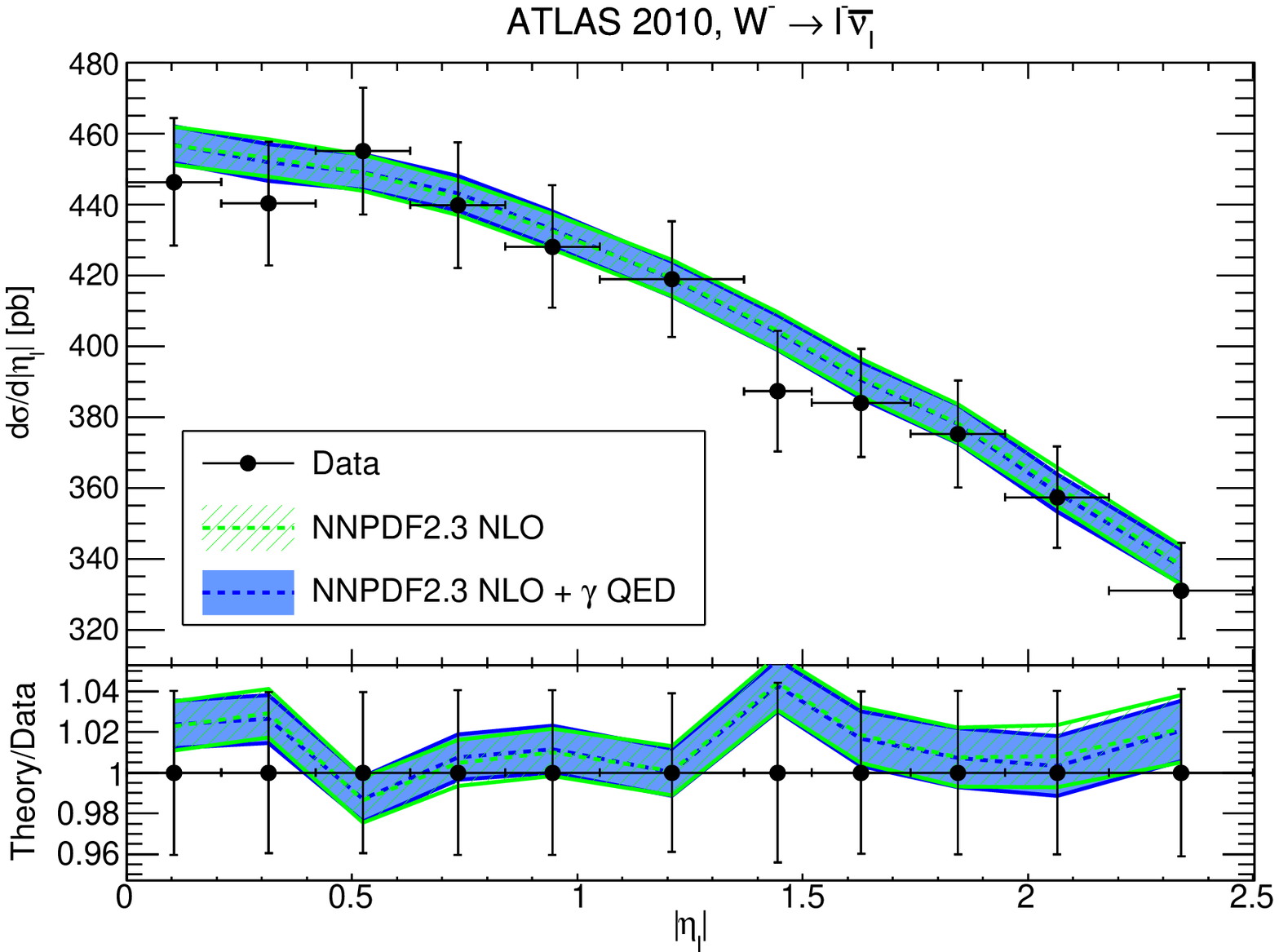}
    \includegraphics[scale=0.39]{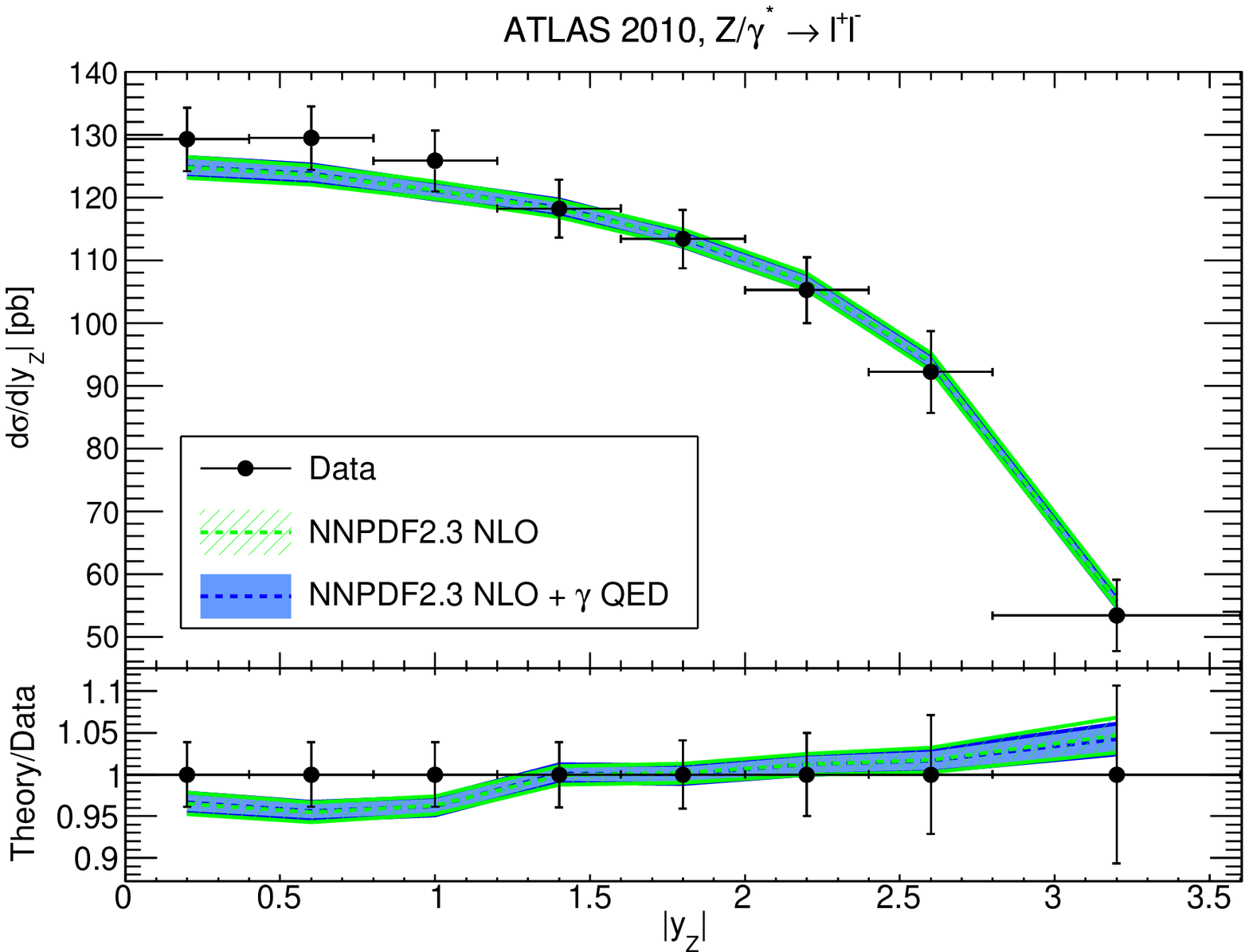}\includegraphics[scale=0.39]{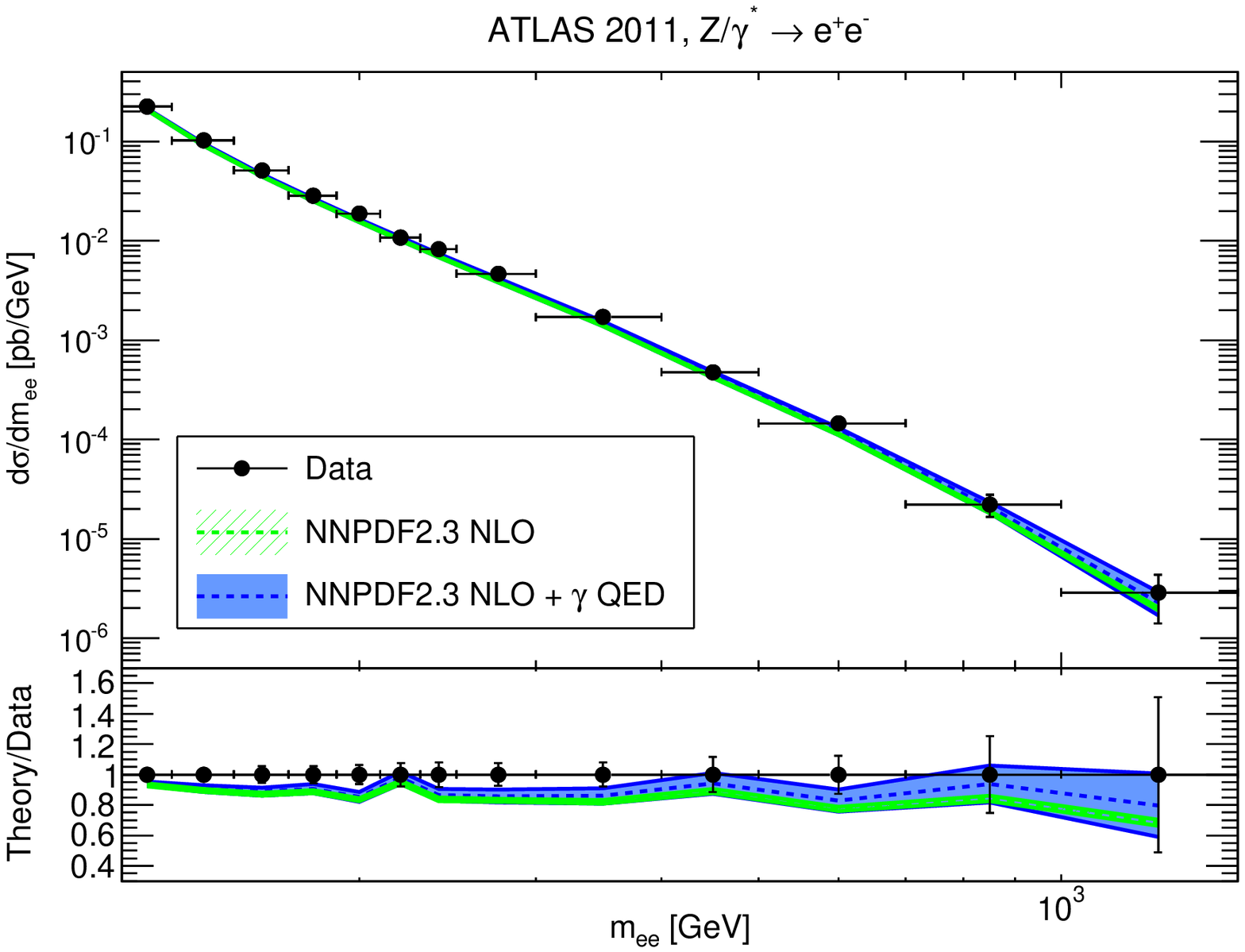}
    \par\end{centering}
  \caption{\label{fig:pheno} Comparison between experimental data and
    theoretical predictions computed with NNPDF2.3 NLO for each data
    set, with (blue lines) and without (green lines) photon-induced
    contributions.}
\end{figure}

Figure~\ref{fig:pheno} shows a comparison between ATLAS data and
theoretical predictions computed with NNPDF2.3 NLO, with and without
photon-induced contributions using the photon PDF obtained after the
reweighting procedure. Again, the comparison shows that the
$Z/\gamma^{*}$ high mass production provides the most relevant
information to constrain the photon PDF.

%%%%%%%%%%%%%%%%%%%%%%%%%%%%%%%%%%%%                                   
%%%%%%%%%%%%%%%%%%%%%%%%%%%%%%%%%%%%

\section{Unweighting the photon PDF}

\begin{figure}
  \begin{centering}
    \includegraphics[scale=0.4]{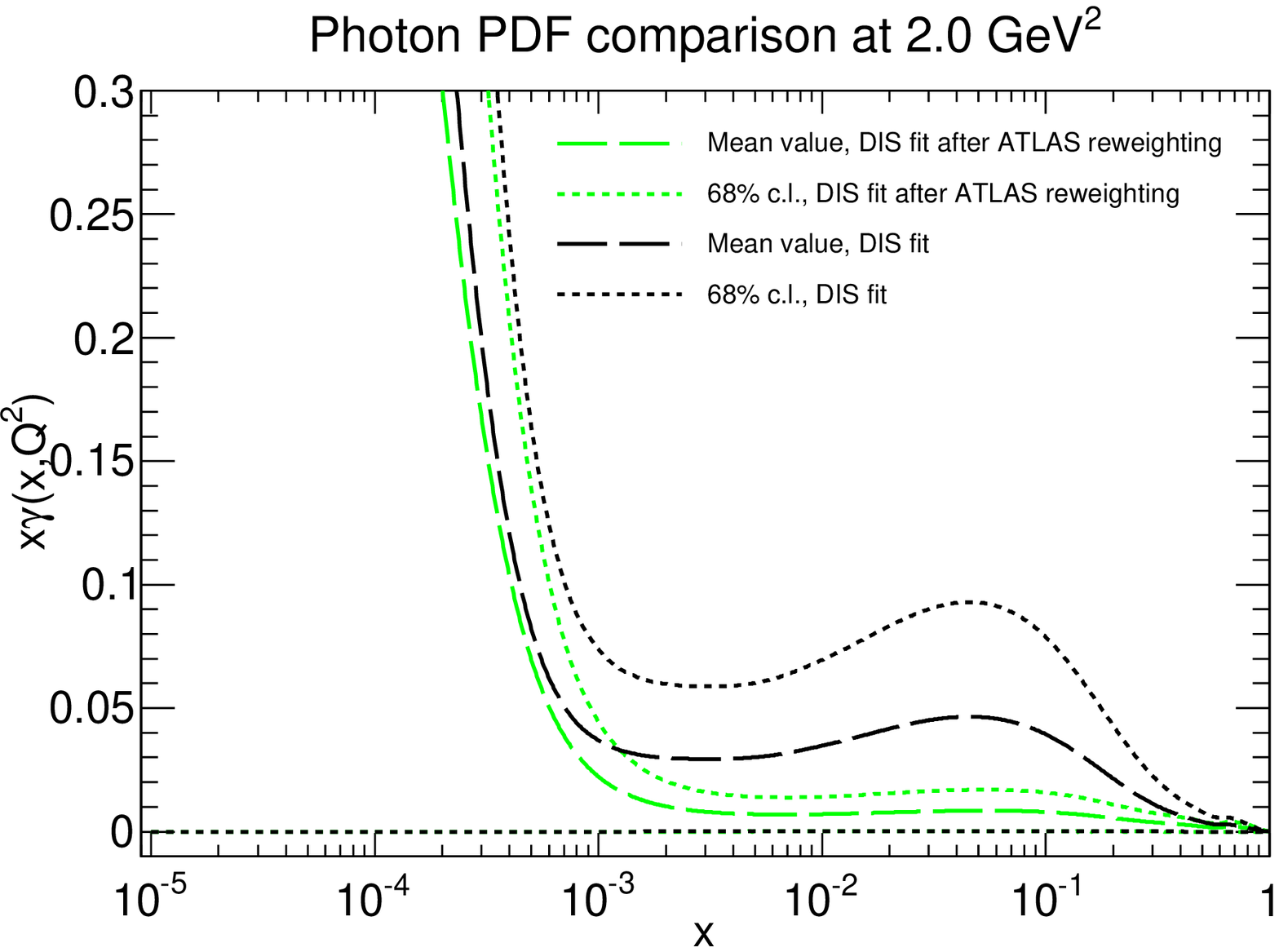}\includegraphics[scale=0.4]{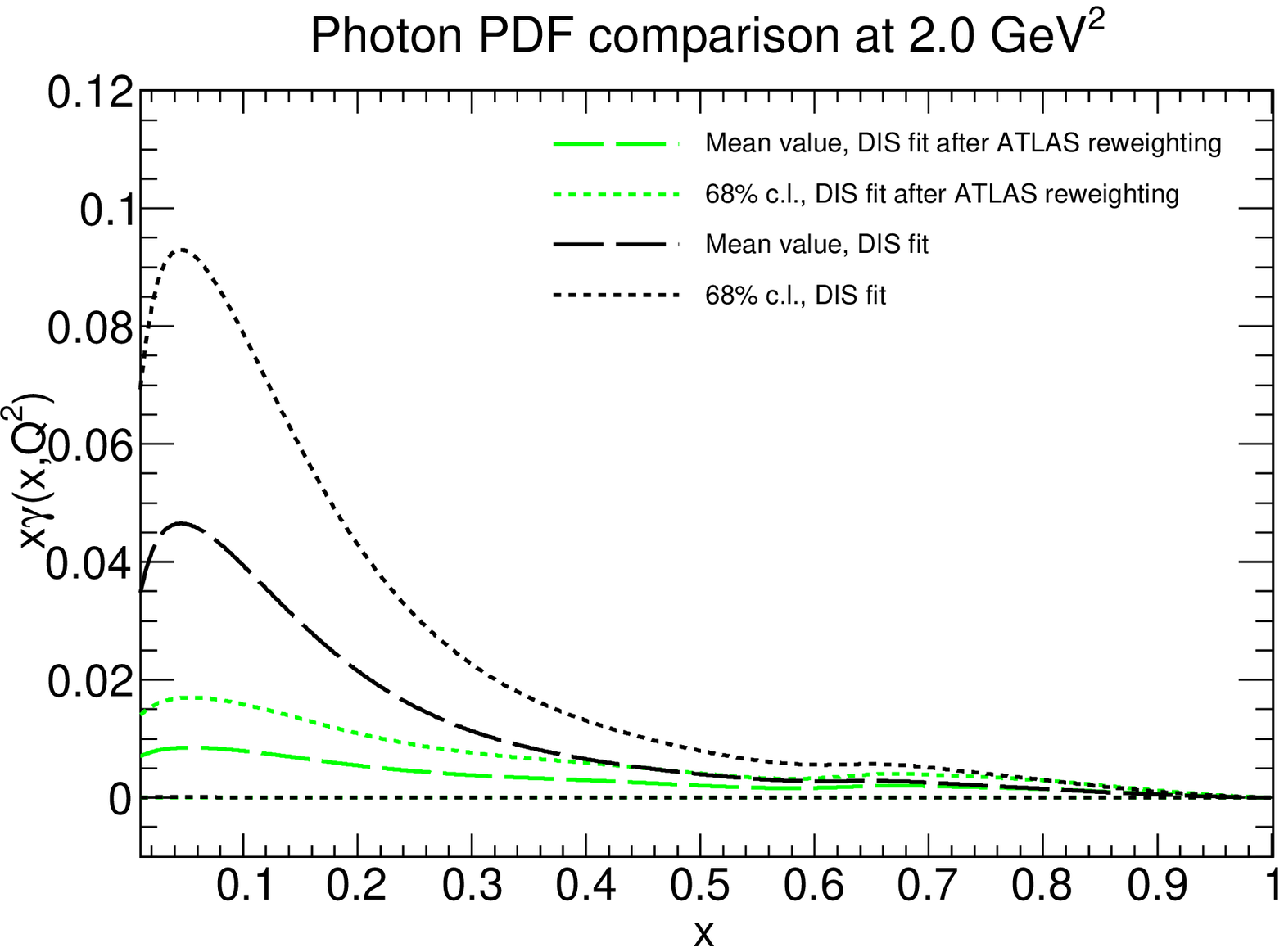}
   \par\end{centering}
 \caption{\label{fig:rw} Direct comparison between the photon PDF
   before (black lines) and after (green lines) the reweighting
   procedure. The reweighted photon PDF uncertainties are reduced at
   medium/small-$x$. The 68\% c.l. band is defined from the PDF mean value.}
\end{figure}

Following the procedure presented in Ref.~\cite{Ball:2011gg} we
construct an unweighted photon PDF set. Figure~\ref{fig:rw} shows the
comparison between the photon PDF before and after performing the
reweighting with ATLAS data, in logarithmic scale (left) and linear
scale (right). We observe a significant reduction of the uncertainties
which essentially overlaps with the coverage in $x$ of the ATLAS data
used in the reweighting. It is interesting to observe that the central
value is also substantially reduced. This means that the central value
from the DIS fit is mostly determined by statistical fluctuations.

\begin{figure}
  \begin{centering}
    \includegraphics[scale=0.4]{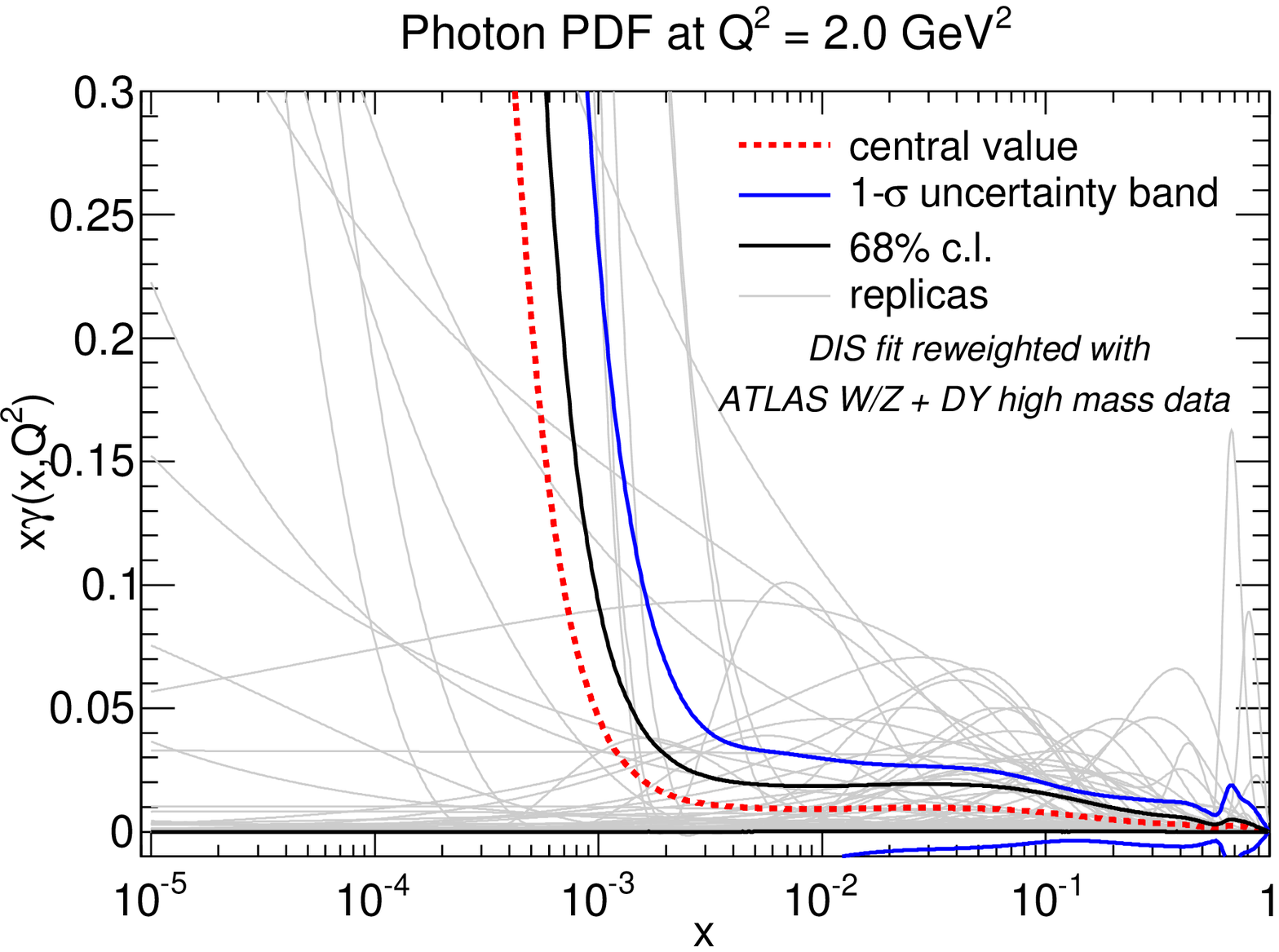}\includegraphics[scale=0.4]{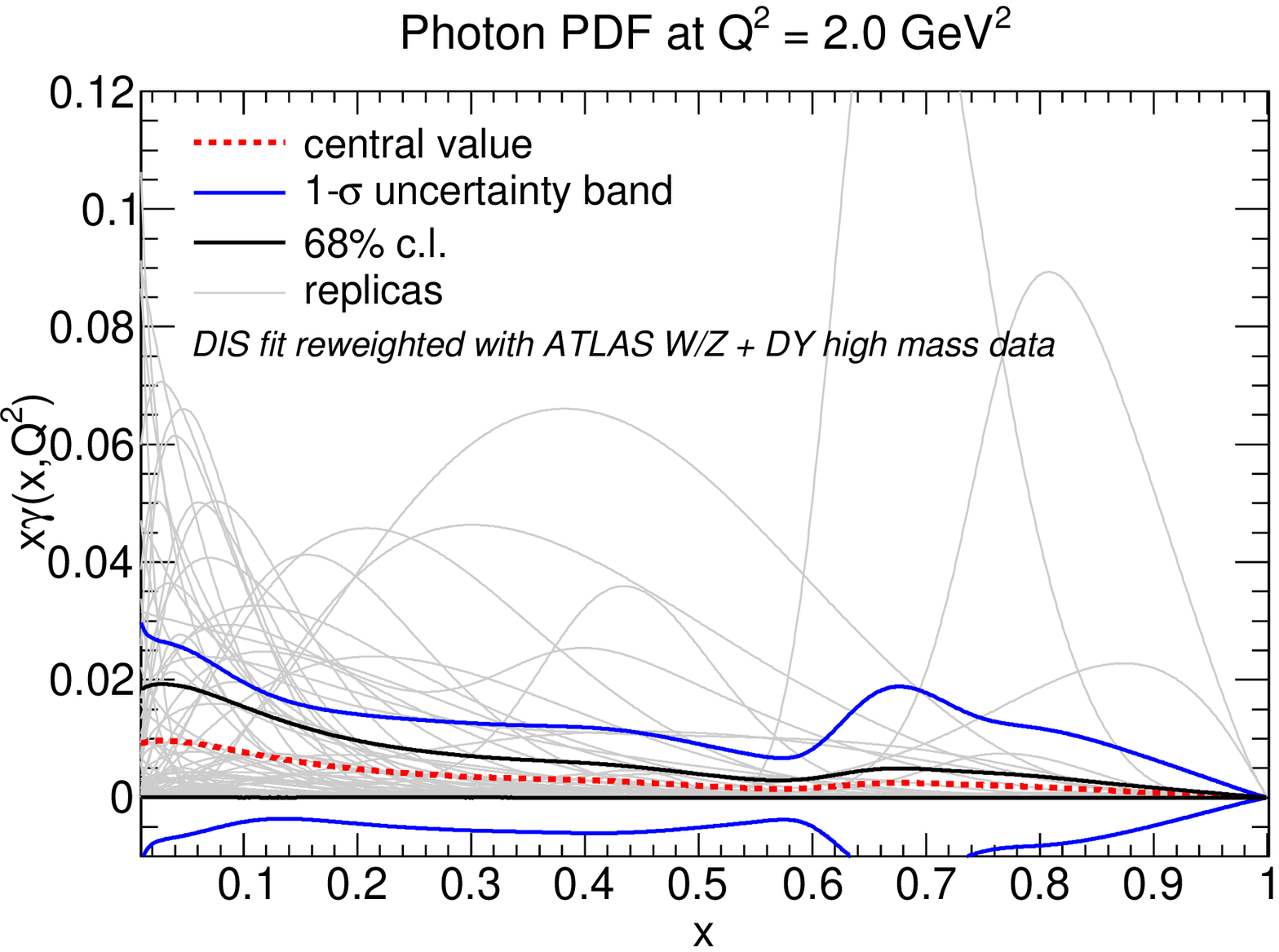}
   \par\end{centering}
 \caption{\label{fig:photon} The unweighted photon PDF at
   $Q^{2}_{0}=2$ GeV$^{2}$ obtained after the reweighting procedure
   with ATLAS $W$, $Z/\gamma^{*}$ rapidity data and
   ATLAS Drell-Yan high mass data. The PDF set includes 100 replicas.}
\end{figure}

Figure~\ref{fig:photon} shows the details of the unweighted photon PDF
set with $N_{\rm rep}=100$ replicas. Also in this example, the current
reweighted photon PDF in comparison to the photon PDF from DIS has
small central values and smaller uncertainties at medium/small-$x$,
improving the quality of theoretical predictions. On the other hand,
we observe that at very small-$x$ the photon PDF has large
uncertainties, because the ATLAS data does not cover this region.

%%%%%%%%%%%%%%%%%%%%%%%%%%%%%%%%%%%%                                   
%%%%%%%%%%%%%%%%%%%%%%%%%%%%%%%%%%%%

%%%%%%%%%%%%%%%%%%%%%%%%%%%%%%%%%%%%                                   
%%%%%%%%%%%%%%%%%%%%%%%%%%%%%%%%%%%%

\section{Outlook}

In conclusion, the Drell-Yan LHC data from ATLAS reduces the photon
PDF uncertainties at medium/small-$x$, providing a much more
constrained and reliable photon PDF in comparison to the results
obtained by fitting DIS-only data.

Here we have shown only the impact of ATLAS data to the photon PDF. In
a the final photon PDF set~\cite{Carrazza:2013} we will also include
the LHCb low mass data, which reduces uncertainties at very
small-$x$. This will provide the first extraction of a modern photon
PDF with uncertainties and constrained by LHC data. Further
improvements of the photon PDF uncertainties will be possible when
more LHC data will be available, specifically, measurements of $WW$
production which at high energy scales constrains the photon PDF at
large-$x$.


\begin{thebibliography}{99}

%\cite{Carrazza:2013wua}
\bibitem{Carrazza:2013wua}
  S.~Carrazza,
  %``Towards an unbiased determination of parton distributions with QED corrections,''
  arXiv:1305.4179 [hep-ph].
  %%CITATION = ARXIV:1305.4179;%%

%\cite{Ball:2012cx}
\bibitem{Ball:2012cx}
  R.~D.~Ball, V.~Bertone, S.~Carrazza, C.~S.~Deans, L.~Del Debbio, S.~Forte, A.~Guffanti and N.~P.~Hartland {\it et al.},
  %``Parton distributions with LHC data,''
  Nucl.\ Phys.\ B {\bf 867} (2013) 244
  [arXiv:1207.1303 [hep-ph]].
  %%CITATION = ARXIV:1207.1303;%%
  %30 citations counted in INSPIRE as of 09 May 2013


%\cite{CarloniCalame:2007cd}
\bibitem{CarloniCalame:2007cd}
  C.~M.~Carloni Calame, G.~Montagna, O.~Nicrosini and A.~Vicini,
  %``Precision electroweak calculation of the production of a high transverse-momentum lepton pair at hadron colliders,''
  JHEP {\bf 0710} (2007) 109
  [arXiv:0710.1722 [hep-ph]].
  %%CITATION = ARXIV:0710.1722;%%
  %80 citations counted in INSPIRE as of 09 May 2013


%\cite{Ball:2010gb}
\bibitem{Ball:2010gb}
  R.~D.~Ball {\it et al.}  [NNPDF Collaboration],
  %``Reweighting NNPDFs: the W lepton asymmetry,''
  Nucl.\ Phys.\ B {\bf 849} (2011) 112
   [Erratum-ibid.\ B {\bf 854} (2012) 926]
   [Erratum-ibid.\ B {\bf 855} (2012) 927]
  [arXiv:1012.0836 [hep-ph]].
  %%CITATION = ARXIV:1012.0836;%%
  %35 citations counted in INSPIRE as of 19 Jun 2013


%\cite{Ball:2011gg}
\bibitem{Ball:2011gg}
  R.~D.~Ball, V.~Bertone, F.~Cerutti, L.~Del Debbio, S.~Forte, A.~Guffanti, N.~P.~Hartland and J.~I.~Latorre {\it et al.},
  %``Reweighting and Unweighting of Parton Distributions and the LHC W lepton asymmetry data,''
  Nucl.\ Phys.\ B {\bf 855} (2012) 608
  [arXiv:1108.1758 [hep-ph]].
  %%CITATION = ARXIV:1108.1758;%%
  %35 citations counted in INSPIRE as of 19 Jun 2013

%\cite{Aad:2011dm}
\bibitem{Aad:2011dm}
  G.~Aad {\it et al.}  [ATLAS Collaboration],
  %``Measurement of the inclusive $W^\pm$ and Z/gamma cross sections in the electron and muon decay channels in $pp$ collisions at $\sqrt{s}=7$ TeV with the ATLAS detector,''
  Phys.\ Rev.\ D {\bf 85} (2012) 072004
  [arXiv:1109.5141 [hep-ex]].
  %%CITATION = ARXIV:1109.5141;%%
  %105 citations counted in INSPIRE as of 19 Jun 2013

%\cite{Aad:2013iua}
\bibitem{Aad:2013iua}
  G.~Aad {\it et al.}  [ATLAS Collaboration],
  %``Measurement of the high-mass Drell--Yan differential cross-section in pp collisions at $\sqrt{s}$=7 TeV with the ATLAS detector,''
  arXiv:1305.4192 [hep-ex].
  %%CITATION = ARXIV:1305.4192;%%

%\cite{Carli:2010rw}
\bibitem{Carli:2010rw}
  T.~Carli, D.~Clements, A.~Cooper-Sarkar, C.~Gwenlan, G.~P.~Salam, F.~Siegert, P.~Starovoitov and M.~Sutton,
  %``A posteriori inclusion of parton density functions in NLO QCD final-state calculations at hadron colliders: The APPLGRID Project,''
  Eur.\ Phys.\ J.\ C {\bf 66} (2010) 503
  [arXiv:0911.2985 [hep-ph]].
  %%CITATION = ARXIV:0911.2985;%%
  %36 citations counted in INSPIRE as of 26 Jun 2013

%\cite{Catani:2009sm}
\bibitem{Catani:2009sm}
  S.~Catani, L.~Cieri, G.~Ferrera, D.~de Florian and M.~Grazzini,
  %``Vector boson production at hadron colliders: A Fully exclusive QCD calculation at NNLO,''
  Phys.\ Rev.\ Lett.\  {\bf 103} (2009) 082001
  [arXiv:0903.2120 [hep-ph]].
  %%CITATION = ARXIV:0903.2120;%%
  %176 citations counted in INSPIRE as of 25 Jun 2013

\bibitem{Carrazza:2013}
  The NNPDF Collaboration,
  %``in preparation,''
  \textit{in preparation}.

\end{thebibliography}
\end{document}